\documentclass[runningheads]{llncs}
\usepackage{graphicx}
\usepackage{amsfonts} 
\usepackage{amsmath}
\usepackage{cite}
\usepackage{todonotes}
\usepackage{enumitem}
\usepackage{paralist}
\usepackage{float}
\usepackage{booktabs}
\usepackage{amssymb}
\begin{document}

\newcommand{\ccos}[2]{C_{#1#2}}
\newcommand{\cmax}[2][d]{C^\star_{#2#1}}
\newcommand{\acmax}[2][d]{#1^\star_{#2}}
\newcommand{\mean}{\mathop{\textrm{mean}}}
\newcommand{\tauAP}{\tau_{AP}}

\title{A White Box Analysis of ColBERT}
%
%
\author{Thibault Formal\inst{1,2} \and
Benjamin Piwowarski \inst{1} \and
Stéphane Clinchant\inst{2}}
\authorrunning{T. Formal, B. Piwowarski and S. Clinchant}
%
\institute{Sorbonne Université, LIP6, F-75005 Paris, France \\ \email{benjamin.piwowarski@lip6.fr} \and
Naver Labs Europe, Meylan, France \\
\email{firstname.name@naverlabs.com}}
\maketitle              
\begin{abstract}
Transformer-based models are nowadays state-of-the-art in adhoc Information Retrieval, but their behavior is far from being understood. Recent work has claimed that BERT does not satisfy the classical IR axioms. However, we propose to dissect the matching process of ColBERT, through the analysis of term importance and exact/soft matching patterns. Even if the traditional axioms are not formally verified, our analysis reveals that ColBERT
\begin{inparaenum}[(i)]
\item is able to capture a notion of term importance;
\item relies on exact matches for important terms.
\end{inparaenum}


\keywords{Information Retrieval  \and Term Matching \and Transformer \and BERT}
\end{abstract}
\section{Introduction}

Over the last two years, Natural Language Processing has been shaken by the release of large pre-trained language models based on self-attention, like BERT \cite{bert}. 
Ranking models based on BERT are currently state-of-the-art for adhoc IR, and rank first on leaderboards\footnote{\url{https://microsoft.github.io/msmarco/}} of the MSMARCO passage and document (re-)ranking tasks by a large margin \cite{passage_ranking}, as well as on more standard IR datasets such as Robust04\cite{cedr, dai_sigir,t5}. They have excelled where previous neural models had been struggling so far \cite{neural_hype}.
It is thus interesting to understand better what is happening inside those models during ranking, and what phenomena are captured. Some works have been conducted in this direction \cite{RenningsAxiomaticApproachDiagnosing2019,Camara_Axioms}, but focused on whether IR axioms are respected -- or not -- by neural and transformer-based models. In\cite{Camara_Axioms}, BERT has been shown to not fully respect axioms that have proved to be important for standard IR models, such as the axiom stating
that words occurring in more documents are less important (IDF effect). Instead of investigating whether these models behave like standard ones, in this paper, we make a step towards  understanding \emph{how} they manage to improve over traditional models through their specific matching process. 

There exists a wide variety of BERT-based ranking models, as summarized in the recent overview \cite{lin2020IRBertReview}. 
Canonical BERT models are difficult to analyse because they require a thorough analysis of attention mechanisms,
which is a complex task \cite{identifiability_transformers_iclr20}. 
We rather choose to focus on contextual interaction models \cite{cedr, 2020interpretable, colbert}, where query and document are encoded \emph{independently} -- contrary to the usual case \cite{passage_ranking}.
Among such models, ColBERT \cite{colbert} exhibits the best trade-off between effectiveness and efficiency, with performance on par with standard BERT, suggesting that the power of these models comes from learning rich contextual representations, rather than modeling complex matching patterns. 
Moreover, the structure of ColBERT (sum over query terms of some similarity scores) is similar to standard IR models like BM25, and makes the analysis easier, as the contribution for each term is explicit.
%

In this paper, we hence focus on ColBERT, and look at two research questions. In Section \ref{sec:term-importance}, we investigate the link between term importance as computed by standard IR models, and the one computed by (Col)BERT. In Section \ref{sec:exact-match}, we look at how (Col)BERT is dealing with exact and soft matches as this is known to be critical for IR systems.

\section{Experimental setting}

\paragraph{Dataset}

For our analysis, we use the passage retrieval tasks from TREC-DL 2019 and 2020 \cite{trec_2019} ($400$ queries in total).
We consider a re-ranking setting, where for a given query $q$, the model needs to re-rank a set of documents $\mathcal{S}_q$ selected by a first stage ranker. Following the MSMARCO setting, we consider candidates from BM25, and $|\mathcal{S}_q| \leq 1000$. In order to study the model properties, we are interested in \emph{how it attributes scores to each query token, for documents in $\mathcal{S}_q$}. 


\paragraph{ColBERT}

We now introduce the variant of ColBERT\cite{colbert} we used to simplify the analysis -- we checked each time that the drop in performance was minor. In particular, we did not include query/document specific tokens (\texttt{[Q]} and \texttt{[D]}), since these tokens could bias the representation of query/document terms.
Second, while query augmentation has been shown to be beneficial in \cite{colbert, hofstatter2020improving}, we omit this component to avoid analysis of the induced implicit query expansion mechanism.
%
We however keep the compression layer, that projects token representations from the BERT representation space ($d=768$) to the ColBERT representation space ($d=128$). By fine-tuning our model in a similar fashion to \cite{colbert}, we obtain a MRR@10 of $0.343$ on MSMARCO dev set (versus $0.349$). This shows that the above simplifications are negligible performance-wise, and would not invalidate our analysis. In order to understand what is learned during training, we also consider a non fine-tuned version of the model (without projection layer), that relies on the output of a pre-trained BERT model. 


The formal definition of ColBERT, given the BERT embeddings $E_q =(E_{q_i})_i$ for the query $q$ (after WordPiece tokenization) and $E_d=(E_{d_j})_j$ for the document $d$, is given by the following relevance score:
\begin{equation}
\label{eq:colbert}
s(q,d) 
= \sum \limits_{i \in q} \max_{j \in d} \text{cos}(E_{q_i}, E_{d_j}) 
= \sum \limits_{i \in q} \max_{j \in d} \ccos ij
= \sum \limits_{i \in q} \cmax i
\end{equation}

In the following, we say that \emph{a query token $i$ matches the document token $j^*$} if $\ccos i{j^*} = \cmax i$. We denote this token $j^*$ by $\acmax i$. 




\section{ColBERT term importance}
\label{sec:term-importance}

Our first research question focuses on comparing the term importance of standard IR models (e.g. BM25) with the term importance as determined by ColBERT. With respect to the former,
given that documents are (small) passages, term frequency is close to $1$ for most terms. Moreover, passage length does not vary much, and is caped at $512$ tokens. Hence, we can reasonably assume that a term BM25 score roughly corresponds to its IDF -- this might not be true for terms with very low IDF values, but it is a good enough approximation for other terms.

For ColBERT, it is difficult to measure the importance of a term because it depends on both document and query contexts.
We hence resort to an indirect mean, by measuring the correlation between the original ColBERT ranking and the ranking obtained when the corresponding word is masked, 
i.e. when we remove from the sum in Equation \eqref{eq:colbert} all the contributions of subwords that compose the word. 
Finally, to compare rankings, we use AP-correlation\footnote{using the Python implementation provided by \cite{tau_ap_python}.} $\tauAP$ \cite{tap}, which is akin to Kendall rank correlation, but gives more importance to the top of the ranking. Values close to $1$ indicate a strong correlation, meaning that the two rankings are similar, implying a low contribution of the term in the ranking process. Note that such measure of importance is query dependent: when the term appears in several queries, we consider the average as a final measure of importance.



\begin{figure}
    \centering
    \includegraphics[width=.7\textwidth]{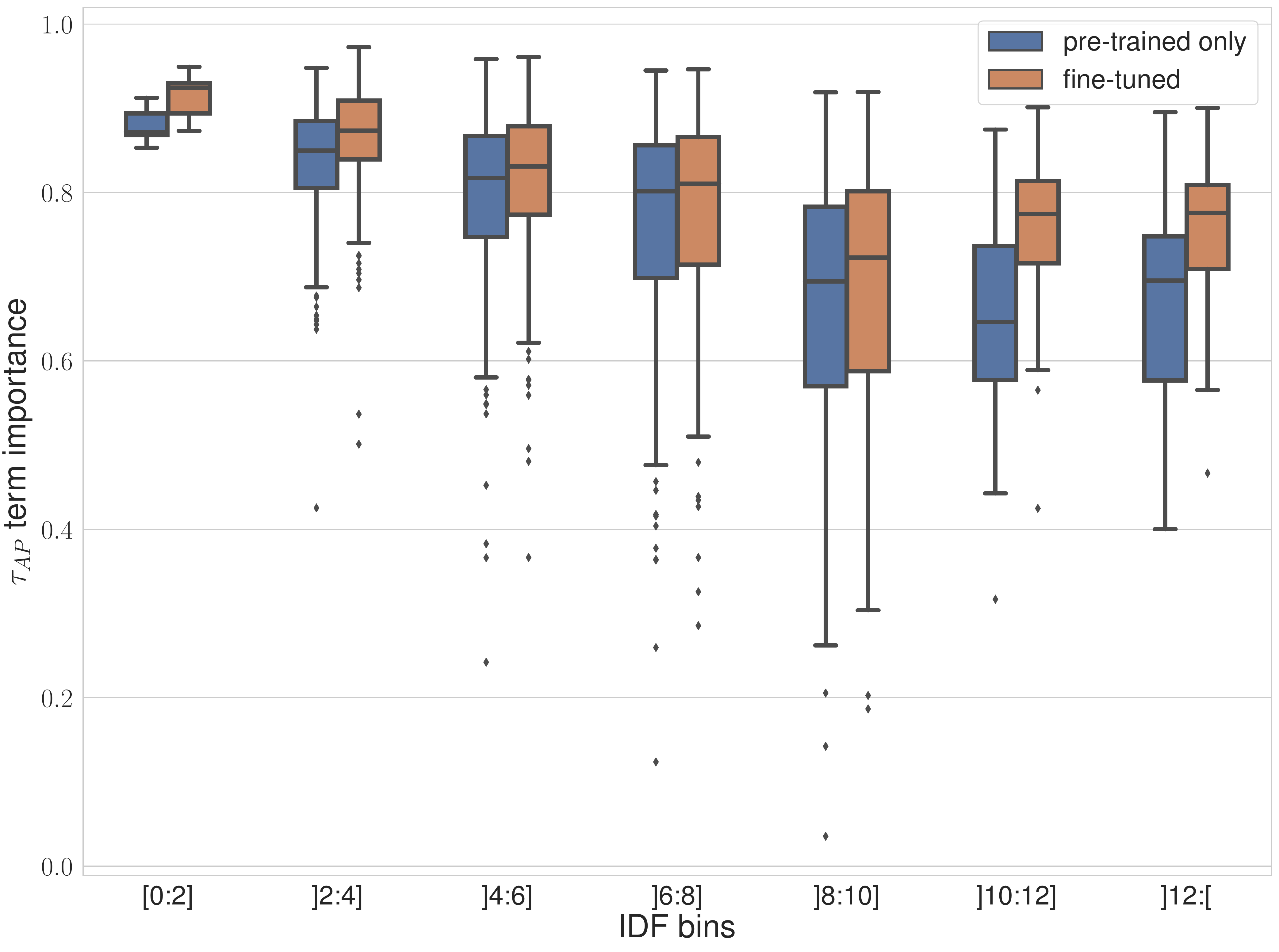}

    \caption{ColBERT term importance (as computed using $\tauAP$) with respect to IDF (standard term importance).}
    \label{fig:idf_vs_tauap}
\end{figure}

In Figure \ref{fig:idf_vs_tauap}, we show how IDF and $\tauAP$ are connected. There is a linear negative correlation between both metrics (Pearson correlation coefficient $r=-0.4$), showing that (Col)BERT implicitly captures IDF. Note that words with higher IDF tend to be longer, and hence to be split into multiple subwords more often -- increasing the importance of such terms. 

We also observe that the link between IDF and term importance is  not so direct for high IDF values ($> 8$). We believe that there are three reasons explaining this behavior:
\begin{inparaenum}[(i)]
\item ColBERT has correctly learnt that this term was not \emph{so} important;
\item as most of the documents contain the term, the effect on $\tauAP$ might not be high;
\item \label{enum:tauap:same} another query term (with no semantics) is bearing the same semantics as the target one. 
\end{inparaenum}

The first hypothesis is probably true since ColBERT improves over BM25. As for the second one, this is a more general observation regarding the re-ranking setting, where IR axioms might not fully apply.
Finally, to investigate the hypothesis (\ref{enum:tauap:same}), we looked, for each query token, at the frequency of exact matching (i.e. the max similarity is obtained with the same token in a document) and at the frequency with which it matches in documents \emph{other terms of the query}. We observed that stopwords (\emph{the}, \emph{of}, etc.) did indeed match terms in the documents that were other query terms. For instance, in the query (and associated $\tauAP$) ``\emph{the (0.94) symptoms (0.87) of (0.93) shingles (0.88)}'', the word ``of'' actually mostly matches with ``shingles'' in documents from $\mathcal{S}_q$. 


\section{Analysis of Exact and Soft matches}
\label{sec:exact-match}

\begin{figure}
    \centering
    \includegraphics[width=.85\textwidth]{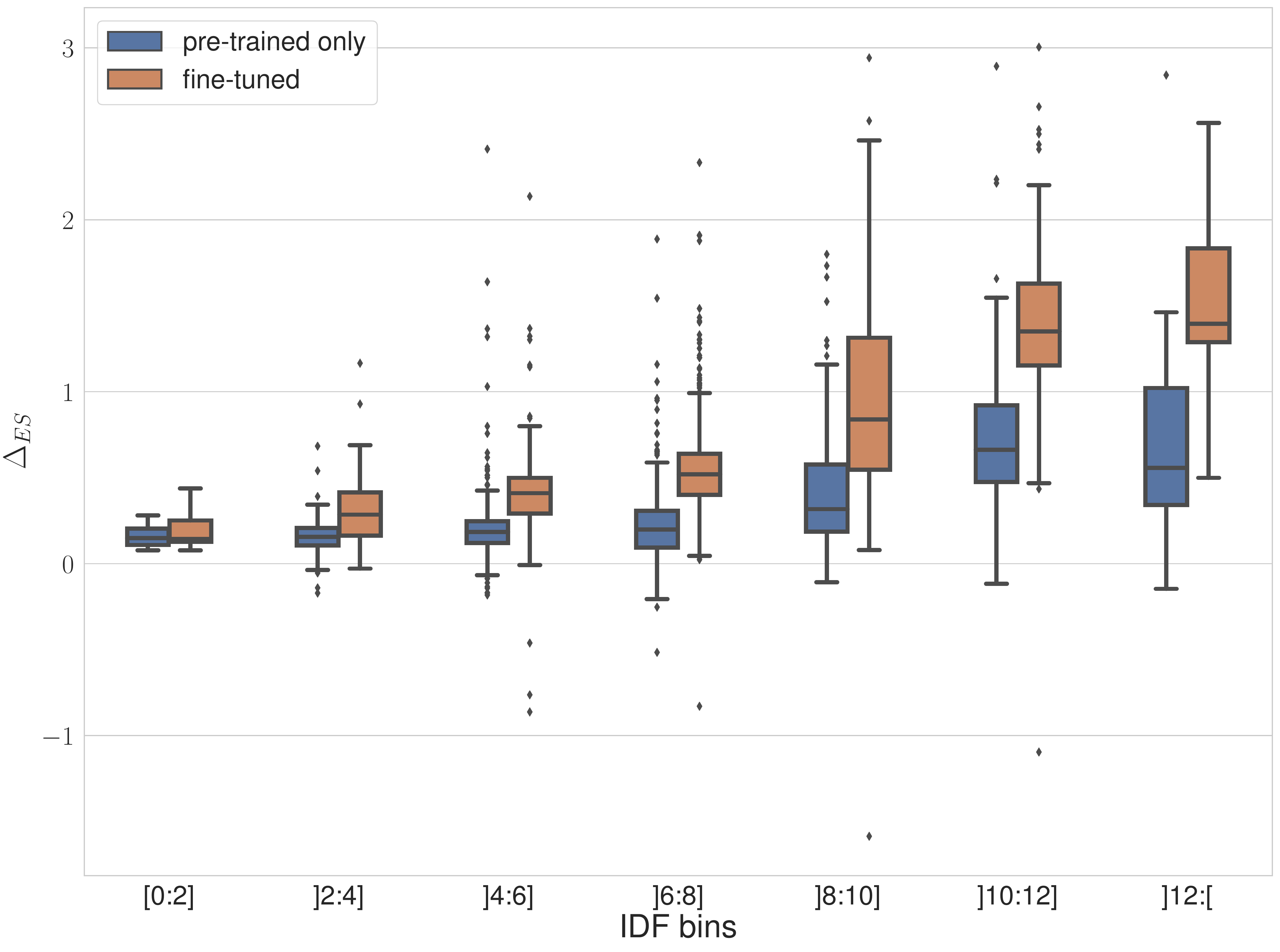}
    \caption{ \small $\Delta_{ES}$ with respect to IDF: we observe a moderate correlation (0.667) between $\Delta_{ES}$ and IDF, showing that the less frequent a term is, the more it is likely to be matched exactly.}
    \label{fig:deltaes_vs_idf}
\end{figure}

After having looked at term importance, we now turn our attention into the issue of exact matches, i.e. how exact string matching is processed by ColBERT. Because it has been trained to re-order a standard term-based IR model, it is interesting to check whether it might be less sensitive to such signals.

To look into this, we need to define a measure indicating when ColBERT asserts whether a term should be an exact match or not (i.e., soft match). To do so, we compute, for each query term $i$, the difference between the average ColBERT scores when $i$ matches the same term within a document (i.e, when $\acmax i \rightarrow t$) or not (i.e., when $\acmax i \not\rightarrow t$). We then average at the query level, to obtain one measure per term (for terms appearing in several queries). This measure is formally defined as:
\begin{equation}
\label{eq:deltaes}
\Delta_{ES} (t) = \mean_{i, q / i \rightarrow t}
    \left( 
        \mean_{d \in \mathcal{S}_q / \acmax i \rightarrow t}\left\{ \cmax i  \right\} \
        - \mean_{d \in \mathcal{S}_q / \acmax i \not\rightarrow t} \left\{   \cmax i  \right\} 
    \right)
\end{equation}
where $j \rightarrow t$ means that the $j^{th}$ token corresponds to token $t$.

For a word $w$ composed of several WordPiece components $t_1, \ldots, t_n$, we use $\sum_{t \in w} \Delta_{ES}(t)$, which corresponds to the way ColBERT works (summing over subwords). 
Then, for all query words $w$, we plot $\Delta_{ES}(w)$ with respect to $IDF(w)$ (Figure \ref{fig:deltaes_vs_idf}). We can observe that there is a moderate positive correlation between terms focusing more on exact matching by ColBERT --larger $\Delta_{ES}$-- and IDF ($r=0.667$). Interestingly, this effect is already observable for BERT, but fine-tuning has an important impact for words with an IDF above $8$: ColBERT thus learns to favor exact matches for such words. For instance, in the query (and associated $\Delta_{ES}$) ``\emph{causes (0.35) of (0.11) left (0.64) ventricular (1.14) hypertrophy (1.62)}'', we can see that the model relies a lot on exact match for the last two terms. \\



\begin{figure}
    \centering
    \includegraphics[width=.9\textwidth]{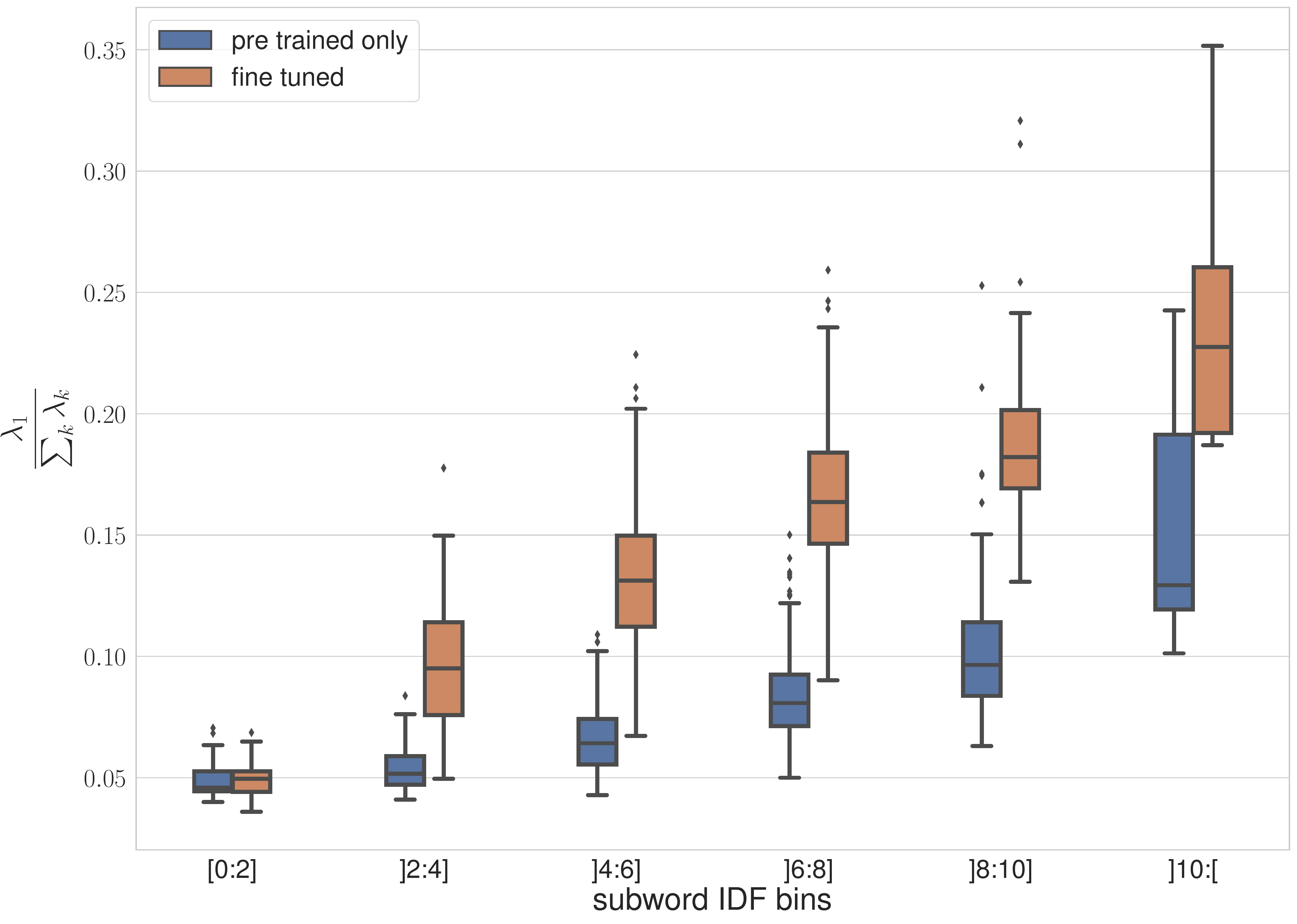}
    \caption{ \small Ratio of the first eigenvalue to the sum of the eigenvalues with respect to IDF (subword level). The less frequent the term is, the higher the ratio is, \emph{showing that all contextualized embedding for a rare term are concentrated in the same direction}.}
    \label{fig:spectral}
\end{figure}

To explain this behavior, our hypothesis is that exact matches correspond to contextual embeddings that do not vary much, while terms that carry less "information" are more heavily influenced by their context (they act as some sort of reservoirs to encode concepts of the sequence), and thus their embeddings vary a lot. To check this hypothesis, we conducted a spectral analysis of contextual term embeddings. More specifically, we use an SVD decomposition of the matrix composed of all the contextual representations for a given term $t$, on the test documents, and look at the relative magnitude of the singular values $\lambda_1 \ge ... \ge \lambda_d$ where $d$ is the dimension of the embedding space. If the magnitude of $\lambda_1$ is much larger than the others, it means that all the contextual representations point to the same direction in the embedding space. 
In Figure \ref{fig:spectral}, we report the ratio of the first eigenvalue $\lambda_1$ with respect to $\sum_k \lambda_k$ for terms that appear in the test queries. This figure confirms the above hypothesis, as the ratio increases with the subword IDF (correlation $r=0.77$). Moreover, this effect is much stronger when fine-tuning ColBERT, indicating that training on relevance indeed promotes exact matches. By looking at the distribution of singular values (not shown here), we can confirm this trend. In particular, words with a low IDF tend to point each time in a different direction, showing that what they capture is more about their context. For instance, in the query ``\emph{when did family feud come out ?}'' (a TV show), the term ``come'', for all the documents in $S_q$, matches $97\%$ of the time to document terms that are not in the query, but are synonyms (in a broad sense) e.g. \emph{\{july, happen, item, landing, released, name, en, going, it, rodgers\}}. 

\section{Conclusion}
While the axiomatic approach is appropriate to analyze traditional IR models, its application to BERT-based models remains limited and somehow inadequate.
To the best of our knowledge, our study is one of the first to shed light on some matching behaviors of BERT, through the analysis of a simpler counterpart, ColBERT.  
We showed that 
\begin{inparaenum}[(i)]
\item even if the IDF effect from the axiomatic theory is not enforced, (Col)BERT does have a notion of term importance;
\item exact matching remains an important component of the model, and is amplified after fine-tuning on relevance;
\item our analysis gave some hints on the properties of frequent words which tend to capture the contexts in which they appear.
\end{inparaenum}

Although this work is a first step towards understanding matching properties of BERT in IR, we believe there is much more to uncover by either analyzing a wider range of models, or by extending our analysis of ColBERT to first stage ranking, where retrieval axioms might be more critical. 

\bibliographystyle{splncs04}
\bibliography{bib.bib}
%





\end{document}